\documentclass{article}
\usepackage{spconf,amsmath,amssymb,amsfonts,epsfig,epstopdf,float}

\newtheorem{theorem}{Theorem}
\newtheorem{definition}[theorem]{Definition}

\title{A Stochastic Game Formulation of Energy-efficient Power Control: Equilibrium Utilities and
Practical Strategies} \name{Fran\c cois M\'eriaux, Ma\"el Le Treust,
Samson Lasaulce, Michel Kieffer}
\address{L2S - CNRS - SUPELEC - Univ Paris-Sud\\
    3 rue Joliot-Curie\\
    F-91192 Gif-sur-Yvette, France\\
    \{meriaux,letreust,lasaulce,kieffer\}@lss.supelec.fr}
\begin{document}
\ninept
\maketitle
\begin{abstract}
Frequency non-selective time-selective multiple access channels in
which transmitters can freely choose their power control policy are
considered. The individual objective of the transmitters is to
maximize their averaged energy-efficiency. For this purpose, a
transmitter has to choose a power control policy that is, a sequence
of power levels adapted to the channel variations. This problem can
be formulated as a stochastic game with discounting for which there
exists a theorem characterizing all the equilibrium utilities
(equilibrium utility region). As in its general formulation, this
theorem relies on global channel state information (CSI), it is
shown that some points of the utility region can be reached with
individual CSI. Interestingly, time-sharing based solutions, which
are usually considered for centralized policies, appear to be part of
the equilibrium solutions. This analysis is illustrated by numerical
results providing further insights to the problem under
investigation.

\end{abstract}
\section{Introduction}

In the past decade, new types of wireless networks have appeared.
Just to name a few, when inspecting the wireless literature, we find
ad hoc networks, networks of wireless devices operating in
unlicensed bands, wireless networks with cognitive radios, small
cells based cellular networks. The decentralized nature of such
networks make game-theoretic analyses relevant~\cite{Mackenzie-book-2006}. Interestingly, game theory offers a large
set of tools and concepts to better understand important problems
such as resources allocation and power control in distributed
networks. By distributed, it is meant that the allocation or/and
control policy is left to the terminal itself. In this paper, the
problem under consideration is precisely the one of distributed
power control. The assumed network model is a multiple access
channel (MAC), which, by definition, includes several transmitters
and one common receiver. A brief overview of previous works about power allocation for MACs is presented in~\cite{belmega-twc-2009}. In our framework, based on a certain knowledge which
includes his individual channel state information, each transmitter
has to tune his power level for each data block. In particular, it can ignore some specified centralized policies. The
assumed performance criterion is the one introduced by~\cite{goodman-pc-2000}
that is to say, that transmitters aim at maximizing the energy
efficiency associated with the transmit radio-frequency signal,
which is measured as a number of correctly decoded information bits
per Joule consumed at the transmitter. The assumed channel model is
the same as~\cite{goodman-pc-2000} that is, links are assumed to be frequency
non-selective but the channel gains can vary from block to block.
The authors of~\cite{goodman-pc-2000} have formalized the energy-efficient power
control problem as a one-shot/static non-cooperative game: on each
block, the players (the transmitters) play a one-shot game by
choosing their action/move (their power level) in order to maximize
their utility (their individual energy-efficiency). The main
drawback of such a formulation is that it leads to an outcome (Nash
equilibrium) which is not efficient. Indeed, it can be checked that,
for each block, there exists a vector of power levels (an action
profile) which allows all the players to have better utilities than
those obtained at the Nash equilibrium; the latter is said to be
Pareto-dominated.

Motivated by the existence of power profiles which Pareto-dominates
the one-shot game Nash equilibrium solution, several authors
proposed solutions which are both efficient and compatible with the
framework of decentralized decisions. For instance, pricing is
proposed in~\cite{saraydar-com-2002}, a Stackelberg formulation is introduced in
\cite{lasaulce-twc-2009}, and a repeated game formulation is exploited in~\cite{LeTreustLasaulce(PowerControlRG)10}. The framework adopted in this paper is a more
general framework than the one chosen in~\cite{LeTreustLasaulce(PowerControlRG)10}.
Indeed, although the repeated game model in~\cite{LeTreustLasaulce(PowerControlRG)10}
takes into account the fact that transmitters interact several/many
times, the work in~\cite{LeTreustLasaulce(PowerControlRG)10} has an important weakness:
this is the need for a normalized game which does not depend on the
realization channels. By definition (see e.g.,
\cite{Sorin92}), a repeated game consists in repeating the
same one-shot game. The consequence of such a modeling choice is a
loss in terms of optimality in terms of expected utilities (averaged
over the channel realizations). The main purpose of this paper is to
propose a more general dynamic game model namely, a stochastic game.
Based on this choice, the contributions of this paper are
essentially as follows: (i) after providing the signal model (Sec.~\ref{sec:signal model}) and reviewing
  the one-shot game model (Sec.~\ref{sec:static game}) we present the stochastic game
  model used and show how to exploit the recent game-theoretic results by~\cite{HornerSugayaTakahashiVieille09}\cite{FudenbergYamamoto09} to obtain a Folk theorem for the power
  control game under investigation: this Folk theorem allows one to
  fully characterize equilibrium utilities when public signals are
  available to the transmitters and channel stats are quantized (Sec.~\ref{sec: stoch games}); (ii) some simple equilibrium power control strategies relying
on individual
    channel state information and possibly a recommendation of the
    receiver are proposed (Sec.~\ref{sec:stoch strat}); (iii) a numerical study is conducted to assess the performance
of
  the proposed strategies and give more insights on tuning the
  relevant parameters of the problem (Sec.~\ref{sec:Numerical}).

\section{Signal Model}
\label{sec:signal model}

We consider a decentralized MAC with $K\geq 1$ transmitters. The
network is said to be decentralized as the receiver (e.g., a base
station) does not dictate to the transmitters (e.g., mobile
stations) their power control policy. Rather, all the transmitters
choose their policy by themselves and want to selfishly maximize
their energy-efficiency; in particular, they can ignore some
specified centralized policies. We assume that the users transmit
their data over quasi-static channels, at the same time and
frequency band and without any beamforming~\cite{Veen-88}. Note that
a block is defined as a sequence of $M\geq1$ consecutive symbols
which contains a training sequence: a specific symbols sequence used
to estimate the channel (or other related quantities) associated
with a given block. A block has therefore a duration less than the
channel coherence time. The signal model used corresponds to the
information-theoretic channel model used for studying MAC,
see~\cite{belmega-twc-2009}
 for more comments on the multiple access
technique involved. This model is both simple
to be presented
 and captures the different aspects of the
problem. It can be readily
applied to specific systems such as CDMA systems \cite{goodman-pc-2000}\cite{lasaulce-twc-2009}
 or multi-carrier CDMA systems \cite{meshkati-jsac-2006}.
 The equivalent baseband signal
received by the base station can be written as
\begin{equation}
 y(n) = \sum_{i=1}^{K} g_i(n) x_i(n) + z(n)
\end{equation}
where $i  \in \mathcal{K}$, $\mathcal{K} = \{1,...,K\}$, $x_i(n)$ represents
the symbol transmitted by transmitter $i$ at time $n$,
$\mathbb{E}|x_i|^2 = p_i$, the noise $z$ is assumed to be
distributed according to a zero-mean Gaussian random variable with
variance $\sigma^2$ and each channel gain $g_i$ varies over time but
is assumed to be constant over each block. For each transmitter $i$,
the channel gain modulus is assumed to lie in a compact set
$\left|g_i \right|^2 \in \left[\eta_i^{\mathrm{min}},
\eta_i^{\mathrm{max}} \right]$. This assumption models the finite receiver
sensitivity and the existence of a minimum distance between the
transmitter and receiver. 
At last, the receiver is assumed to implement single-user decoding.

 At a given instant,
 the SINR at receiver $i \in \mathcal{K}$ writes as:
\begin{equation}
\mathrm{SINR}_i=\frac{p_i \eta_i}{\sum_{j \neq i} p_j
\eta_j+\sigma^2}
\end{equation}
where $p_i$ is the power level for transmitter $i$ and $\eta_i =
|g_i|^2$.

.

\section{One-shot power control game}
\label{sec:static game}

In this section, the one-shot game model of~\cite{goodman-pc-2000} is reviewed
since it both allows one to build the stochastic game model of Sec.~\ref{sec: stoch games} and serves as a reference for performance comparison. A useful
(non-equilibrium) operating point in this game is also defined, as a
basis for the proposed control strategies in the stochastic game
model.

\begin{definition}[One-shot power control game] The strategic form of the one-shot
power control
game is a triplet
\newline
$\mathcal{G} =
(\mathcal{K},
\{\mathcal{A}_i\}_{i\in\mathcal{K}},\{u_i\}_{i\in\mathcal{K}})$
where $\mathcal{K}$ is the set of players,
$\mathcal{A}_1,...,\mathcal{A}_K$ are the corresponding sets of
actions, $\mathcal{A}_i = [0, P_i^{\mathrm{max}}]$,
$P_i^{\mathrm{max}}$ is the maximum transmit power for player $i$,
and $u_1,...,u_K$ are the utilities of the different players which
are defined by:
\begin{equation}
\label{eq:def-of-utility} u_i(p_1,...,p_K)= \frac{R_i
f(\mathrm{SINR}_i)}{p_i} \ [\mathrm{bit} / \mathrm{J}].
\end{equation}
\end{definition}

We denote by $R_i$ the transmission information rate (in bps) for user $i$ and $f$ an efficiency function representing the block success rate. The numerator of the utility is thus the rate of bits successfully received at the base station. $f$ is assumed to be sigmoidal and identical for all the users; the sigmoidness assumption is a reasonable assumption, which is well justified in \cite{rodriguez-globecom-2003}\cite{meshkati-tcom-2005}. Recently, \cite{belmega-valuetools-2009} has shown that this assumption is also justified from an information-theoretic standpoint.

In this game with complete information ($\mathcal{G}$ is known to every
player) and rational players (every player does the best for himself
and knows the others do so and so on), a major game solution
concept is the Nash equilibrium (i.e. a point from which no player has interest
in unilaterally deviating). When it exists, the non-saturated\footnote{By using the term ``non-saturated Nash equilibrium'' we mean that
the maximum transmit power for each user, denoted by $
P_i^{\mathrm{max}}$, is assumed to be sufficiently high not to be
reached at the equilibrium i.e. each user maximizes his
energy-efficiency for a value less than $P_i^{\mathrm{max}}$ (see
\cite{lasaulce-twc-2009} for more details about the saturated case).} Nash equilibrium of
this game can be obtained by setting $\frac{\partial u_i}{\partial
p_i}$ to zero $\forall i \in \mathcal{K}$ which gives an equivalent condition on the SINR: the
best SINR in terms of energy-efficiency for transmitter $i$ has to
be a solution of $xf'(x)-f(x)=0$ (this solution is independent of
the player index since a common efficiency function is assumed, see
\cite{meshkati-tcom-2005} for more details). This leads to:
\begin{equation}
 \forall i \in \{1,...,K \}, \
p_i^{*}= \frac{\sigma^2}{\eta_i} \frac{\beta^*}{1-(K-1)\beta^{*}}
\label{eq:NE-power}
\end{equation}
where $\beta^{*}$ is the unique solution of the equation
$xf'(x)-f(x)=0$. An important property of
the Nash equilibrium given by (\ref{eq:NE-power})
is that transmitters only need
to know their individual channel gain $\eta_i$ to play their
equilibrium strategy. Another interesting property is that the
product $p_{i}^* \eta_i$ (instantaneous received power) is constant for all the players. 

Interestingly, the authors of~\cite{LeTreustLasaulce(PowerControlRG)10} propose to study
the power profile obtained when imposing the received signals to
have the same instantaneous power. The idea is to solve
$\frac{\partial u_i}{\partial p_i}(\underline{p}) = 0$ under the
aforementioned constraint. This leads to the following system:
\begin{equation}
\label{eq:def-operating-point}
\forall (i,j) \in \mathcal{K}^2, \left\{\begin{array}{ll}
p_i \eta_i=p_j \eta_j \\
\frac{\partial u_i}{\partial p_i}(\underline{p}) = 0
\end{array}
\right.
\end{equation}

The unique solution of
(\ref{eq:def-operating-point}), called \textit{Operating point} can be checked to be:
\begin{equation}
\label{eq:power-profile-op-pt} \forall i \in \mathcal{K}, \
\tilde{p_i} =
\frac{\sigma^2}{\eta_i}\frac{\tilde{\gamma}_K}{1-(K-1)\tilde{\gamma}_K}
\end{equation}
where $\tilde{\gamma}_K$ is the unique solution of $ x[1-(K-1)\cdot x]
f'(x)-f(x)=0$.

The difference between this operating point and the Nash equilibrium can be explained by the fact that $\frac{\partial u_i}{\partial p_i}(\underline{p})=\frac{\partial u_i}{\partial p_i}(p_i)$ when adding the instantaneous power equality constraint. This operating point can be proved to always Pareto-dominate the
Nash equilibrium and reach the Pareto frontier for each channel
realization; additionally, only individual CSI is needed to operate
at the corresponding power levels. This point will serve as a basis
of the power control strategies proposed in Sec.~\ref{sec:stoch strat}.

\section{Stochastic power control game}
\label{sec: stoch games}

With the one-shot non-cooperative game model, transmitters are
assumed to play once for each bock and independently from block to
block. The goal here is to take into account the fact that
transmitters generally interact over several blocks, which is likely
to change their behavior w.r.t. the one-shot interaction model even
if they are always assumed to be selfish. As channel gains are
time-varying, the most natural model is the one of stochastic
games~\cite{Shapley53}. In such a model, important differences
w.r.t. the one-shot game model are that averaged utilities are
considered, the channel state $\underline{\eta}(t) = (\eta_1(t),
..., \eta_K(t)) \in \Gamma$ and $\Gamma = \Gamma_1 \times \Gamma_2
\times... \times \Gamma_K $ may vary according to a certain
evolution law (the i.i.d. block fading case is the most simple of
them), and the state can depend on the played actions (in
conventional wireless settings this is however not the case).
Stochastic game stages correspond to instants at which players can
choose their actions. From one stage to another, the channel state
$\underline{\eta}(t)$ is assumed to be discrete (e.g., resulting
from quantization effects) and stochastically varies according to
the transition probability distribution $\pi$. This distribution is
said to be an irreducible transition probability if for any pair of
channel states $\underline{\eta}$ and $\underline{\eta}'$ we have
$\pi(\underline{\eta}'|\underline{\eta})>0$. For example, this
irreducibility condition is met for i.i.d channels. The second
important assumption we do to obtain a Folk theorem is to assume
that a public signal is available to all the transmitters. Two
special cases of interest are: (a) Every transmitter knows the power
of the received signal that is, $\sum_{i=1}^K \eta_i p_i + \sigma^2$
is known; (b) Every transmitter has global CSI and perfectly
observes the action profiles that is, $(\underline{\eta},
\underline{p})$ is known.

\textit{The game Course.} The game starts at the first stage with a
channel state $\underline{\eta}(1)$ known by the players. The
transmitters simultaneously choose their power level
$\underline{p}(1)=(p_1(1),\ldots,p_K(1))$ and are assumed to receive
a public signal, denoted $\theta \in \Theta$. At stage $t$, the
channel states $\underline{\eta}(t)$ are drawn from the transition
probability $\pi(\cdot|\underline{\eta}(t-1))\in \Delta(\Gamma)$ and
the players observe the public signal
\begin{equation}
\phi: \left|
\begin{array}{ccc}
\Gamma \times \mathcal{A} & \rightarrow & \Theta\\
(\underline{\eta}, \underline{p}) & \mapsto & \theta
\end{array}
\right.
\end{equation}
where $\mathcal{A} =\mathcal{A}_1 \times ...\times\mathcal{A}_K$.
The sequence of past signals $\underline{h}(t) =
(\theta(1)...,\theta(t-1), \underline{\eta}(t))$ is the common
history of the players.
\begin{definition}[Players' strategies]
A strategy for
player $i \in \mathcal{K}$ is a sequence of functions $\left(\tau_{i,t}
\right)_{t \geq 1}$ with
\begin{equation}
\tau_{i,t}: \left|
\begin{array}{ccc}
\Theta^t & \rightarrow & \mathcal{A}_i \\
 \underline{h}_t & \mapsto & p_i(t).
\end{array}
\right.
\end{equation}
\end{definition}

The strategy of player $i$ will therefore be denoted by $\tau_i$
while the vector of strategies $\underline{\tau} = (\tau_1, ...,
\tau_K)$ will be referred to a joint strategy. A joint strategy
$\underline{\tau}$ induces in a natural way a unique sequence of
action plans $(\underline{p}(t))_{t\geq 1}$ and a unique sequence of
public signals $(\underline{\theta}(t))_{t\geq1}$.
 The averaged utility for player $i$ is defined as follows.
\begin{definition}[Players' utilities]
Let $\underline{\tau} = (\tau_1, ..., \tau_K)$ be a joint strategy.
The utility for player $i \in \mathcal{K}$ if the initial channel
state is $\underline{\eta}(1)$, is defined by:
\begin{equation}
v_i(\underline{\tau}, \underline{\eta}(1)) = \sum_{t
\geq 1} \lambda (1 - \lambda)^{t-1}
\mathbb{E}_{\underline{\tau},\pi}\left[u_i(\underline{p}(t),
\underline{\eta}(t))|\underline{\eta}(1)\right]
\end{equation}
where $(\underline{p}(t))_{t \geq 1}$ is the sequence of power
profiles induced by the joint strategy $\underline{\tau}$.
\end{definition}
The parameter $\lambda$ is the discount factor, which can model
various effects such as the probability the game stops, the fact
that players evaluate short-term and long-term benefits differently,
etc. We now present the definition of a stochastic game.
\begin{definition}[Stochastic game]
\label{def:stoch game}
A stochastic game with public monitoring is defined as a tuple:
\begin{equation}
\mathcal{G}=(\mathcal{K},(\mathcal{T})_{i\in\mathcal{K}},
(v_i)_{i\in\mathcal{K}},(\Gamma_i)_{i\in\mathcal{K}},\pi, \Theta, \phi),
\end{equation}
where $\mathcal{T}_i$ is the set of strategies for player $i$ and
$v_i$ his long-term utility function.
\end{definition}

\emph{Equilibrium concept.} Let us define the Nash equilibrium of a
 stochastic game starting with the channel state $\underline{\eta}(1)$.
\begin{definition}[Equilibrium Strategies]
A strategy $\underline{\tau}$ supports an equilibrium of the
stochastic game with initial channel state $\underline{\eta}(1)$ if
\begin{equation} \forall i \in \mathcal{K}, \forall \tau_i', \
v_i(\underline{\tau},\underline{\eta}(1)) \geq v_i(\tau_i',
\underline{\tau}_{-i},\underline{\eta}(1))
\end{equation}
where $-i$ is the standard notation to refer to the set $\mathcal{K}
\backslash \{i\}$, \\ $\underline{\tau}_{-i} = (\tau_1, ...,
\tau_{i-1}, \tau_{i+1},...,\tau_K)$. Denote $E_{\lambda}(\eta(1))$
the set of equilibrium utilities with initial channel state
$\underline{\eta}(1)$.
\end{definition}

We now characterize the equilibrium utility region for the
stochastic game under study. For this, define the min-max level
$\tilde{v}_i$ of player $i\in \mathcal{K}$ as the most severe
punishment level for player $i\in \mathcal{K}$. The feasible utility
region is denoted by $F_{\lambda}(\underline{\eta}(1))$. The result
of Dutta \cite{Dutta95} states that if the transition probability
$\pi$ is irreducible, then the min-max levels, the feasible utility
region and the equilibrium utility region are independent of the
initial state $\underline{\eta}(1)$.
\begin{eqnarray}
\lim_{\lambda \longrightarrow 0} \min_{\tau_{-i}}\max_{\tau_{i}}
\tilde{v}_i(\tilde{\tau}_i,\underline{\tilde{\tau}}_{-i},\underline{\eta}(1))&=&\tilde{v}_i
,\; \forall  \underline{\eta}(1)\\
\lim_{\lambda \longrightarrow 0} F_{\lambda}(\underline{\eta}(1))&=&F,\; \forall \underline{\eta}(1)\\
\lim_{\lambda \longrightarrow 0} E_{\lambda}(\underline{\eta}(1))&=&E,\; \forall \underline{\eta}(1)
\end{eqnarray}
As already mentioned, irreducibility is verified under the common
assumption of i.i.d. channels. As a consequence, the equilibrium
utility region is independent of the initial channel state for a
stochastic game with a public signal.
\begin{definition} We define the set of asymptotic feasible and individually rational
payoff by: \begin{equation}F^*=\{x \in F| x_i\geq \tilde{v}_i,\;
\forall i \in \mathcal{K}\}\end{equation}
\end{definition}
\begin{theorem} Suppose that the players see the same public signal. Then, for
 each utility vector $\underline{u} \in F^*$,
there exists
 a $\lambda_0$ such that for all $\lambda < \lambda_0$, there
  exists a perfect public equilibrium strategy in the
  stochastic power control game, such that the
  long-term utility equals $\underline{u}\in F^*$.
\end{theorem}
The proof is not given here and is based on Theorem 2 of
\cite{HornerSugayaTakahashiVieille09}. Note that such a
characterization is very powerful. Indeed, the brute-force technique
to find the feasible utility region would be to look at all possible
action plans for the players and compute the corresponding
utilities. This would be intractable even when every player could
only choose two power levels for a finite stochastic games with $100$
stages (each player could then choose between $2^{100}$ action plans).
The Folk theorem characterizes the equilibrium utilities from
quantities
 far much easier to evaluate (namely $E,F, \tilde{v}_i$).
 Additionally, as shown by Dutta~\cite{Dutta95}, there is no loss of
 optimality by restricting the set of strategies to Markov
 strategies (which only depend on the current channel state). This
 will be exploited in Sec.~\ref{sec:Numerical}

\section{Strategies for K-player games}
\label{sec:stoch strat}

\subsection{Best user selection (BUS)}

\label{sec:best combi} The strategy we propose is based on the
operating point presented in Sec.~\ref{sec:static game}. When channels gains vary from
stage to stage, if every
 transmitter plays at the operating
point, the network is not socially optimal (in contrast with the
case where the channel state would be constant). It turns out that
we get better results in terms of social welfare if the set of
players playing the operating point at each stage is shrunk. We
name this approach the \textit{best user selection} scheme.

At each stage $t$ of the game, the receiver sets $\mathcal{K}^{'t}
\subset \mathcal{K}$, the optimal set of players playing the
\textit{Operating point} in terms of social welfare. For player $i
\in \mathcal{K}$: $\bullet$ If $i \in \mathcal{K}^{'t}$, he is
recommended to play the \textit{Operating point} at stage $t$;
$\bullet$ If $i \notin \mathcal{K}^{'t}$, he should not transmit at
this stage. To ensure the equilibrium of this strategy, if one of
the player deviates from the strategy, the other players punish him
by playing the
 one-shot \textit{Nash equilibrium} for the remaining of the game.

\subsection{Threshold-based user selection (T-US)}
\label{sec:threshold}

Note that in the former strategy, the set of players playing the
\textit{Operating point} is decided at each stage but one can also
imagine a simpler strategy with a threshold $\alpha \in [0,1]$ set
for the entire game such that for player $i \in \mathcal{K}$:
$\bullet$ If $\eta_i^t\, \geq\,\alpha\, \eta_{\max}^t$, he is
recommended to play the \textit{Operating point}; $\bullet$ If
$\eta_i^t\, <\,\alpha\, \eta_{\max}^t$, he should not transmit at
this stage, where $\eta_{\max}^t$ is the best channel gain
realization at stage $t$ and $\eta_i^t$ the channel gain realization
of player $i$ at stage $t$. As before, if one of the players
deviates from the strategy, the other players punish him by playing
the one-shot \textit{Nash equilibrium} for the remaining of the
game.

\subsection{Properties of Best User Selection Scheme}

Although one might think that the BUS scheme
 requires complex computations at the base receiver, it happens
  that the former does not have to compare
  all possible combinations of players. Indeed, the
  BUS scheme always includes the player with the
  best channel gain and a given number of other players with
  the following channel gains in decreasing order. Thus, the base
  station just has to order the channels gains in decreasing
  order and decides which gain level is the minimum to separate
  players allowed to play from the others, which is far
  less complex, especially when the number of players is very
  large. In term of complexity, for $K$ players, instead of
  considering $2^K$ possible combinations, the base station has just to consider $K$ combinations.

\begin{theorem}[BUS scheme for $k$ players]
\label{th:best combination} At equal transmitting rate, the best
user selection of $k$ players, $k\in \mathcal{K}$, playing together at the
Operating point is the set of the $k$ players with the best channels
gains.
\end{theorem}

For the strategy to be an equilibrium of the stochastic game, it is needed that no player has interest in deviating from the given plan. This condition is expressed by the fact that the cost of the punishment must always be higher than what a player can get when deviating at one stage. It results in the following theorem
\begin{theorem}[Equilibrium strategy]
\label{th:strat_eq} The BUS strategy is an equilibrium of the
stochastic game if $\ \forall i \in \mathcal{K}$
\begin{equation}
\lambda \leq  \frac{\mathbb{E}[u_i(\underline{p}^{bus},\underline{\eta})]-\mathbb{E}[u_i(\underline{p}^*,\underline{\eta})]}{\frac{R\eta_{\max}}{\sigma^2}\frac{f(\beta^*)}{\beta^*}+\mathbb{E}[u_i(\underline{p}^{bus},\underline{\eta})]-\mathbb{E}[u_i(\underline{p}^*,\underline{\eta})]}
\end{equation}
\end{theorem}

The expected utility of the BUS scheme can be
compared to the strategy based on the one-shot \textit{Nash equilibrium}, the
strategy based on pure \textit{Time-sharing}\footnote{At each stage, only the player with the best channel gain can play.} and the \textit{Operating
point} strategy.
\begin{theorem}[Dominance]
\label{th:domi} For equal transmission rates, for i.i.d. channel states among the players,
we have
\[
\forall i \in \mathcal{K},\,
\mathbb{E}[u_i(\underline{p}^{bus},\underline{h})]\, \geq\, \mathbb{E}[u_i(\underline{p}^*,\underline{h})]
\]
\[
\forall i \in \mathcal{K},\,
\mathbb{E}[u_i(\underline{p}^{bus},\underline{h})]\, \geq\, \mathbb{E}[u_i(\tilde{\underline{p}},\underline{h})]
\]
\[
\forall i \in \mathcal{K},\,
\mathbb{E}[u_i(\underline{p}^{bus},\underline{h})]\, \geq\, \mathbb{E}[u_i(\underline{p}^{ts},\underline{h})]
\]
\end{theorem}

Simulations based on this strategy are discussed in Sec~\ref{sec:sim
rayleigh}.

\subsection{Information assumptions}

For the BUS and the TUS schemes, players adapt their transmit power at the
\textit{Operating point} if they are recommended to play by the Base
station. Thus, they need to know whether they are recommended to
play. If so, given (\ref{eq:def-operating-point}), they need to know
the number of players transmitting with them and the state of their
own channel. The recommendation and the number of recommended
players are sent by the Base station whereas the channel state is observed by the
transmitter, all these signals are modeled by $\theta$. The last necessary piece of
side information is due to the equilibrium condition of the
strategy: if one player deviates from the plan, he is punished by
the other players for the remaining of the game. It implies that
players must be able to detect a deviation, what can be done if they
know their SINR at each stage
(see~\cite{LeTreustLasaulce(PowerControlRG)10} for more details).
From one strategy to another, the amount of information required may
vary considerably. For instance, if one wants players to play the
\textit{Social Optimum} strategy, players would have to know
channels gains of every other player whereas if they play the
one-shot \textit{Nash equilibrium} they would only need to know
their own channel gain and the total number of players. To have a
clear view about the amount of side information needed to implement
the discussed strategies, we provide a simplified comparison for
the various strategies under study in Fig.~\ref{fig:side info}. By
``deviation alarm'' we mean a signal allowing players to detect
single deviations from the cooperative action plan.

\begin{figure*}
\vspace{-1.5cm}
\centering
\begin{tabular}{|c|c|c|c|c|}
\hline
\,&\textbf{CSI}&\textbf{Recommendation signal}&\textbf{Nb of players}&\textbf{Deviation
 alarm}\\
\hline
\textbf{Pure time-sharing}& individual & needed & not needed & not needed  \\
\hline
\textbf{One-shot Nash}& individual & not needed & needed & not needed  \\
\hline
\textbf{Operating Point}& individual & not needed & needed& SINR \\
\hline
\textbf{T-US scheme}& individual & needed & needed & SINR \\
\hline
\textbf{BUS scheme}& individual & needed & needed & SINR \\
\hline
\textbf{Social optimum}& global & not needed & not needed & SINR \\
\hline
\end{tabular}
\caption{Information required for various stochastic strategies}
\label{fig:side info}
\end{figure*}

\section{Numerical results}
\label{sec:Numerical}
\subsection{Simulation Parameters}
To obtain numerical results, we work with the efficiency function $f(\gamma)\,=\,e^{-\frac{a}{\gamma}}$ with $a = 2^R-1$, see~\cite{belmega-twc-2009} for more details about this efficiency function. All our results are obtained from games with $10^5$ stages.



\subsection{Two-state channel $K$-player game}
\label{sec:sim two-state}

\begin{figure}[H]
\includegraphics[scale=0.45]{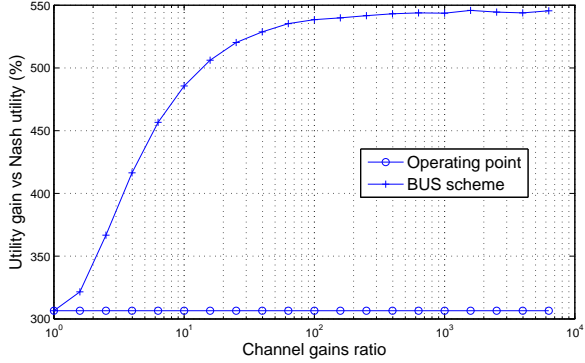}
\caption{Comparison of BUS utilities versus Nash utilities depending on the ratio between the good channel state and the bad channel state for a $10$-player game.}
\label{Fig:two-player two states}
\end{figure}

Fig.~\ref{Fig:two-player two states} is obtained for a $10$-player game where channels gains can only reach two states $\eta_{\min}$ and $\eta_{\max}$ with probability $\{\frac12,\frac12\}$ for each player and $a=0.1$. It is interesting in the sense that it clearly highlights the fact that the idea of the proposed strategy is interesting when channels gains of players are sufficiently different. Indeed, we can observe that for a channel gain ratio $\frac{\eta_{\max}}{\eta_{\min}} = 1$, the proposed strategy is equal to the classical \textit{Operating point} strategy. For $\frac{\eta_{\max}}{\eta_{\min}}>1$, the proposed strategy becomes more efficient. By working with random distributions for the channels gains, the case of realizations with very different channels gains often occurs, which is interesting for the proposed strategy.

\begin{figure}[H]
\includegraphics[scale=0.5]{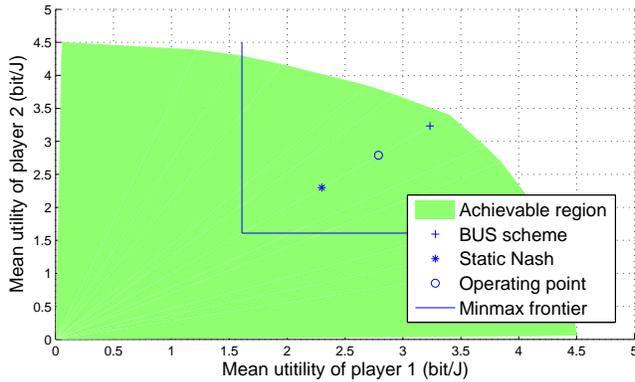}
\caption{Achievable region with expected utilities of various strategies}
\label{Fig:Folk}
\end{figure}

Fig.~\ref{Fig:Folk} shows the achievable utility region for a $2$-player, $2$-states game (with $\frac{\eta_{\max}}{\eta_{\min}}=4$ and $a=0.5$) when considering all the possible strategies. The \textit{minmax} line delimits the equilibrium region. The mean utilities of BUS, \textit{Operating point} and static \textit{Nash equilibrium} are also represented in this region. It is clear that BUS strategy is closer to the Pareto frontier.

\subsection{Truncated Rayleigh distribution for $K$-player game}
\label{sec:sim rayleigh} 
\begin{figure}
\hspace{-1cm}
\includegraphics[scale=0.45]{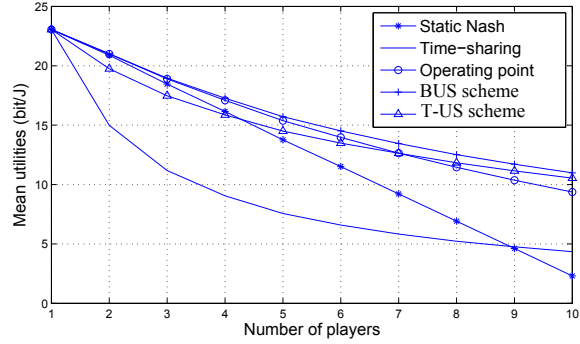}
\caption{Comparison of strategies utilities depending on the number of players}
\label{Fig:uti rayleigh}
\end{figure}

In Fig.~\ref{Fig:uti rayleigh}, we
consider a $K$-player game with $K$ varying from $1$ to $10$ and $a$ fixed to $0.1$. In
each game, channels states follow the same truncated Rayleigh
distribution law for every player. Four strategies are compared on
this figure: one-shot \textit{Nash equilibrium} (Sec.~\ref{sec:static
game}), pure \textit{Time-sharing}, \textit{Operating point}
(equation~(\ref{eq:power-profile-op-pt})), T-US with $\alpha = 0.5$ (Sec.~\ref{sec:threshold}) and BUS (Sec.~\ref{sec:best combi}). There are several points to
notice. First, for all the studied strategies, as the number of
players increases, the mean utility decreases for each player. This
is due to the fact that players see each other signal as
interference: the more players in the game, the more they have to share resources.
Second, it is clear that the \textit{Operating point} and
BUS strategies are more efficient than
one-shot \textit{Nash equilibrium} or \textit{Time-sharing}
strategies. As the number of players increases, this gap becomes
even larger.



\begin{figure}[H]
\includegraphics[scale=0.5]{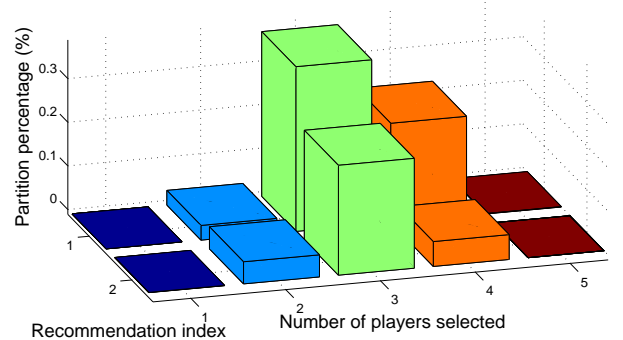}
\caption{Partition of $\mathcal{H}_1^i$ and $\mathcal{H}_2^i$ for a $5$-player game with BUS strategy}
\label{Fig:partition}
\end{figure}

Fig.~\ref{Fig:partition} shows a graphic representation of the different configurations $\mathcal{H}_1^i(k)$ and $\mathcal{H}_2^i(k)$ a player $i$ can meet at each stage of the stochastic game (owing to the symmetry in the channels distribution law, the configurations probabilities are the same for every player). $\mathcal{H}_1^i(k) \subset \Gamma$ is the set of channels realizations where $k$ players are recommended to play and player $i$ is part of these players. $\mathcal{H}_2^i(k) \subset \Gamma$ is the set of channels realizations where player $i$ is not one of he $k$ players recommended to play. The simulation is made with $5$ players and $a=0.2$.

\begin{figure}[H]
\includegraphics[scale=0.5]{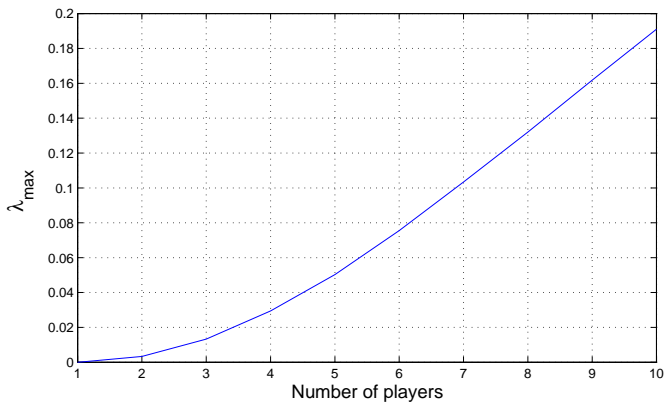}
\caption{Maximum value of discount factor $\lambda$ for the BUS strategy to be an equilibrium.}
\label{fig:lambda_max}
\end{figure}

Fig.~\ref{fig:lambda_max} refers to theorem~\ref{th:strat_eq}. It
represents the maximum value the discount factor $\lambda$ can have
for the BUS strategy to be an equilibrium of the
stochastic game. As the gap between the expected utility of Best
combination strategy and the expected utility of the one-shot Nash
equilibrium increases with the number of players, the maximum value
of the discount factor increases as well.

\section{Conclusion}

Conventionally, for i.i.d channels, power control schemes are
designed such that the power levels are chosen in an independent
manner from block to block. In distributed networks with selfish
transmitters, the point of view has to be re-considered even if the
channels are i.i.d. due to the fact that long-term interaction may
change the behavior of selfish transmitters. In order to take into
account this effect and the fact that channel gains may vary from
block to block, the model of stochastic games is proposed. When
transmitters observe a public signal (e.g., the sum of received
signals), a recent game-theoretic result allows one to fully
characterize the equilibrium utility region. It is shown how to
reach some points of this region by assuming individual CSI only.
Both analytical and simulation results show potential gains in terms
of energy-efficiency induced by the proposed model. In particular,
because of long-term interaction, transmitters may have interest of
shutting down for some blocks, leading therefore to legitimate
time-sharing based control policies. Further investigations on the
proposed approach are needed. In particular, it is relevant to
characterize which part of the equilibrium utility region can be
reached under the individual CSI assumption, which is relevant to
include fairness issues. Additionally, typical features of modern
wireless networks such as finite size buffers, Markovian evolution
law for the channel state, should be accounted for to make the
proposed framework more applicable.

\bibliographystyle{IEEEbib}
\bibliography{BiblioMael}

\end{document}